\documentclass[aps,preprint,groupedaddress]{revtex4}%PRE
\usepackage{amsfonts}
\usepackage{mathrsfs}
\usepackage{graphicx}
\usepackage{amsmath}
\usepackage{amssymb}
\usepackage{amsbsy}
\usepackage{bm}
\usepackage{soul}
\usepackage{xcolor}

\newcommand*{\bb}[1]{\mathbf{#1}}
\newcommand*{\reff}[1]{Eq.~(\ref{#1})}
\DeclareMathOperator\dif{d\!}
\DeclareMathOperator\Dif{D\!}

\newcommand\mi{\mathrm{i}}

\newcommand\x{\hat{x}}
\newcommand\h{h_{l\to\mu}}
\newcommand\y{y_{l\to\mu}}
\newcommand\xl{x_{l\to\mu}}
\DeclareMathOperator\sgn{sgn}

\begin{document}
    
\title{Equivalence between algorithmic instability and transition to replica symmetry breaking in perceptron learning systems}
\author{Yang Zhao}
\author{Junbin Qiu}
\author{Mingshan Xie}
\author{Haiping Huang}
\email{huanghp7@mail.sysu.edu.cn}
\affiliation{PMI Lab, School of Physics,
Sun Yat-sen University, Guangzhou 510275, People's Republic of China}

\date{\today}
\begin{abstract}
    Binary perceptron is a fundamental model of supervised learning for the non-convex optimization, which is a root of the popular deep learning.
    Binary perceptron is able to achieve a classification of random high-dimensional data by computing the marginal probabilities of binary synapses.
    The relationship between the algorithmic instability and the equilibrium analysis of the model remains elusive. Here, we establish the relationship
    by showing that the instability condition around the algorithmic
    fixed point is identical to the instability for breaking the replica symmetric saddle point solution of the free energy function.
    Therefore, our analysis would hopefully provide insights towards other learning systems in bridging the gap between non-convex learning dynamics and statistical mechanics properties of more
    complex neural networks.
\end{abstract}

\maketitle
\section{Introduction}
Theoretical studies of neural networks become increasingly important in recent years~\cite{SMDL-2020,Lenka-2020,HH-2022}, as deep neural networks are widely used
in various domains of both scientific and industrial communities. One of the most powerful theoretical
tools is the replica method, which is able to derive equilibrium properties of neural networks (systems of interacting neurons or synapses), such as
phase diagram~\cite{Amit-1985,Barkai-1992pra,Barra-2018,Huang-2020}, storage capacity~\cite{Gardner-1987,Gardner-1988a,Gardner-1989,Krauth-1989}, and even large-deviation behavior
of learning algorithms~\cite{Baldassi-2015,Baldassi-2016b,baldassi-2021}. Intuitively, the replica method introduces $n$ (an integer) copies of the original
system. Within each copy, there exist strong interactions among constituent elements (e.g., synaptic or neural states), and these interactions
make the model intractable without any approximation in most cases. However, the elements would become decouple with each other as an overlap (of states) matrix is introduced,
which allows a hierarchical level of approximation depending on the stability analysis of the saddle points of the free energy action. A seminal approximation, namely replica symmetry breaking,
was introduced by Giogio Parisi in 1980s~\cite{Parisi-1979b,Parisi-1980b}. 

Overall, the replica method, despite its non-intuitive physics, could lead to exact results in some models. One drawback of this method is that it could not be used to design any efficient algorithms in neural 
networks. Instead, cavity method is constructed via a physically intuitive way, i.e., a statistical mechanics model of learning can be mapped onto a graphical model, where interactions are represented by
factor nodes, and synapses are represented by variable nodes (such as the graphical model representation in unsupervised learning~\cite{Huang-2017a}). Through virtually deleting these two kinds of nodes, a cavity
probability could be defined. Using the tree-like structures of the factor graph, or weakly-interacting-element assumption, an iterative equation for these cavity probabilities can be derived, which
leads to self-consistent evaluations of thermodynamic quantities, such as ground-state energy, free energy and entropy~\cite{cavity-1989,cavity-2001,cavity-2003}.
Most interestingly, this iterative equation is exactly the same as the belief propagation developed independently in computer science~\cite{Yedidia-2005}. The belief propagation could be also derived for learning problems with discrete synapses
~\cite{Zecchina-2006}. An open question is whether the replica symmetry breaking transition corresponds to the algorithmic instability in learning of neural networks.

Here, we provide a proof about this fundamental equivalence in the seminal model of binary perceptron learning, in which learning is achieved by adjusting discrete synapses (actually, the synaptic state takes $\pm1$).
This model is first studied by Gardner and Derrida~\cite{Gardner-1988,Gardner-1989}. A follow-up calculation showed that the storage capacity of this model is given by
$P_c\simeq0.833N$~\cite{Krauth-1989}, where $N$ is the number of neurons, and $P_c$ is the critical number of random patterns being correctly classified.
The binary perceptron belongs to the NP-hard class in the worst case complexity. The typical weight-configuration is quite hard to find by any algorithms based on local flips (e.g., Monte-Carlo dynamics)~\cite{Horner-1992a,Patel-1993,Huang-2010jstat,Huang-2011epl}.
A first efficient algorithm was inspired by the cavity method, reaching an algorithmic threshold $P_{\rm alg}\simeq0.72N$. It was then proved by defining a distance-dependent potential that the entire solution space
is composed of single valleys of vanishing entropy~\cite{Huang-2014pre}. This picture was further shown mathematically rigorous in some perceptron learning problems~\cite{Sly-2021}.
However, the region of the solution space accessed by practical algorithms does not belong to the equilibrium hard-to-reach isolated parts,
but subdominant dense parts~\cite{Baldassi-2015,Baldassi-2016b}. These dense parts are further shown to have good generalization properties~\cite{Baldassi-2018,baldassi-2021}, providing a new paradigm to understand deep learning.

Therefore, studying the mathematical foundation of the binary percetpron problem is fundamentally important to our understanding of neural networks.
To our best knowledge, there are rare studies on the relationship between the replica symmetry breaking transition and the belief propagation instability along this line.
In this work, we show how the belief propagation instability is connected to the instability of the replica symmetric (RS) solution (or replicon mode) of the model.

\begin{figure}
	\centering
	\includegraphics[bb=7 6 406 281,scale=0.8]{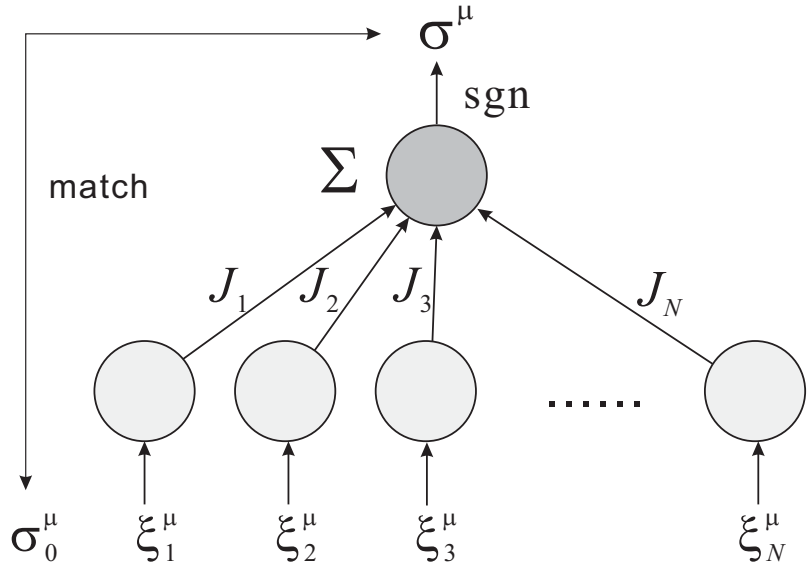}
	\caption{Sketch of a binary perceptron.
	The binary perceptron is composed of $N$ input units (light gray circles) connected to one output unit (dark gray circle). 
	Each input unit receives one pixel of the input pattern. 
	The output unit computes a weighted sum (indicated by $\Sigma$), which is passed through a non-linear sign function (indicated by 
	$\sgn$) to carry out a binary classification.
	The task of the binary perceptron is to find a set of weights that can match all the actual classification ($\{\sigma^\mu\}$) of given patterns with their labels ($\{\sigma_0^\mu\}$).
	}
	\label{bperc}
\end{figure}

\section{Binary perceptron}
Binary perceptron is a single-layer neural network that learns a random input-output mapping by discrete synapses (see Fig.~\ref{bperc}).
We assume that there are $P$ uncorrelated input-output associations, where the $\mu$-th one consists
of an $N$-dimensional pattern $\bm{\xi}^\mu$ and a corresponding label $\sigma^\mu_0$, where $\xi^\mu_i$ and $\sigma^\mu_0$ take $\pm 1$ with equal probabilities. 
Given a configuration of synaptic weights $\{J_i\}_{i=1}^N$ (each entry takes $+1$ or $-1$), the binary perceptron gives the output $\sigma^\mu=\sgn{\left(\sum_{i=1}^NJ_i\xi^\mu_i\right)}$ for the input pattern $\bm{\xi}^\mu$. 
If $\sigma^\mu=\sigma^\mu_0$, we say that the synaptic weight vector $\bb{J}$ has recognized the $\mu$-th pattern. The binary perceptron is able to store an extensive number of random patterns. Therefore,
we define a loading rate $\alpha=P/N$.
When the loading rate is below some threshold, there exists at least a set of synaptic weights as a solution to correctly classify
all the patterns. However, as $\alpha$ exceeds the threshold, it is impossible to find a compatible configuration of weights for all patterns~\cite{Krauth-1989}.
This threshold is also defined as the storage capacity.
Naturally, we define the energy of this model as the number of misclassified patterns as follows,
\begin{equation}
    E(\bb{J})=\sum^P_{\mu=1}\Theta\left(-\frac{\sigma^\mu_0}{\sqrt{N}}\sum_i J_i \xi^\mu_i\right),
\end{equation}
where $\Theta\left(x\right)$ is a step function with the convention that $\Theta\left(x\right)=0$ if $x\leq0$ and $\Theta\left(x\right)=1$ otherwise. 
The prefactor $1/\sqrt{N}$ ensures that the statistical mechanics analysis leads to extensive free energy.

In the zero-temperature limit, the flat measure over the weights realizing the pattern-label associations can be computed as
\begin{equation}\label{bdJ}
    P(\bb{J})=\frac{1}{\mathrm{Z}}\prod_{\mu}\Theta\left(\frac{\sigma^\mu_0}{\sqrt{N}}\sum_i J_i \xi^\mu_i\right),
\end{equation}
where $\mathrm{Z}$ is not only the partition function but also the number of solutions for the learning problem. Equation (\ref{bdJ})
can be derived from the finite temperature Boltzmann measure $P(\bb{J})\propto e^{-\beta E(\bb{J})}$.
Notice that there is a gauge transformation $\xi^\mu_i\to\xi^\mu_i\sigma^\mu_0$ to each pixel of the input patterns that does not affect the Boltzmann measure.
We thus assume $\sigma^\mu_0=+1$ for all patterns in the following analysis. 

\section{Mean-field message passing equations for learning}

The belief propagation (BP) algorithm is an iterative mean-field equation to calculate the marginal probabilities of synaptic state
by passing beliefs between two types of nodes (function nodes and variable nodes)~\cite{HH-2022}. In other words, the beliefs or cavity probabilities can be assumed as messages and thus the BP
algorithm is actually a mean-field message passing equation.
Taking pattern-classification constraints as function nodes and synaptic weights as variable nodes, 
we obtain the iterative equations for learning as follows~\cite{Zecchina-2006,Huang-JPA2013}
\begin{subequations}
\label{bpercbp}
\begin{align}
    m_{i\to \nu}&=\tanh\left(\sum_{\mu\neq \nu}u_{\mu\to i}\right),\\
    u_{\mu\to i}&=\frac12\left[\ln{H\left(-\frac{\frac1{\sqrt{N}}\xi^\mu_i+w_{\mu\to i}}{\sqrt{\sigma_{\mu\to i}}}\right)}-\ln{H\left(-\frac{-\frac1{\sqrt{N}}\xi^\mu_i+w_{\mu\to i}}{\sqrt{\sigma_{\mu\to i}}}\right)}\right],\\ 
    w_{\mu\to i}&=\frac1{\sqrt{N}}\sum_{j\neq i}m_{j\to \mu}\xi^\mu_j,\\
    \sigma_{\mu\to i}&=\frac{1}{N}\sum_{j\neq i}\left(1-m_{j\to \mu}^2\right),
\end{align}
\end{subequations}
where $H(x)=\int_x^\infty{\rm D}z$, ${\rm D}z$ is a Gaussian measure, $m_{i\to \nu}$ is a cavity magnetization parameter to parameterize the cavity probability
$P\left(J_i|\{\bm{\xi}^{\mu\neq \nu}\}\right)=\left(1+m_{i\to \nu}J_i\right)/2$.
$w_{\mu\to i}$ and $\sigma_{\mu\to i}$ represent the mean and variance of the Gaussian distribution of $U_{\mu\to i}\equiv\frac{1}{\sqrt{N}}\sum_{j\neq i}J_{j}\xi_j^\mu$, respectively.
We have applied the centre-limit theorem to the sum $U_{\mu\to i}$ of weakly-correlated terms. This mean-field approximation must be cross-checked by numerical experiments.
In the following analysis, we use
$\mu,\nu$ to indicate function nodes or pattern constraints, and $i,j$ to indicate the variable nodes.

We remark that Eq.~(\ref{bpercbp}) can be combined with an iterative reinforcement to develop an efficient solver. The reinforcement is a kind of soft-decimation, which 
progressively enhances or weakens current local fields (a summation of cavity biases $u_{\mu\to i}$) with an increasing probability with iterations. The algorithm terminates once a solution is found.
This procedure yields the algorithmic threshold $\alpha_{\rm alg}\simeq0.72$~\cite{Zecchina-2006}. During the stochastic reinforcement, the BP iteration does not require convergence, despite convergence guarantee
below the storage capacity. The algorithmic threshold is later found to be below a large-deviation threshold $\alpha_{\rm LD}\simeq0.77$ after which the subdominant dense clusters 
fragment into separate regions~\cite{Baldassi-2015,Baldassi-2016a}. However, our current analysis is restricted to the original BP iteration [Eq.~(\ref{bpercbp})],
rather than the dynamics of reinforced BP and the geometric landscape. It remains challenging to use our framework (without a lengthy replica computation) 
to derive the landscape geometry which relies heavily on replica formula. The following analysis may shed light on this important research line.

\section{Time evolution of message distributions}
In this section, we study the iteration dynamics of the belief propagation.
In the large-$N$ limit, $u_{\mu\to i}$ can be approximated by the first-order Taylor expansion
\begin{equation}
    u_{\mu\to i}=\frac{\xi^\mu_i}{\sqrt{N\sigma_{\mu\to i}}}\frac{G\left(-\frac{w_{\mu\to i}}{\sqrt{\sigma_{\mu\to i}}}\right)}{H\left(-\frac{w_{\mu\to i}}{\sqrt{\sigma_{\mu\to i}}}\right)},
\end{equation}
where $G(x)=\exp(-x^2/2)/\sqrt{2\pi}$, and $H\left(x\right)\equiv\int^\infty_x\Dif z$ with the Gaussian measure $\Dif z\equiv G(z)\dif z$. 
At the iteration step $t$, the macroscopic distributions of messages $m_{l\to\mu}^t$ and $u_{\mu\to l}^t$ are given by:
\begin{subequations}
\begin{align}
    \pi_1^t(x)&=\frac{1}{NP}\sum^N_{l=1}\sum^P_{\mu=1}\delta\left(x-m_{l\to\mu}^t\right),\\
    \pi_2^t(\x)&=\frac{1}{NP}\sum^N_{l=1}\sum^P_{\mu=1}\delta\left(\x-u_{\mu\to l}^t\right).
\end{align}
\end{subequations}
According to the Kabashima's method \cite{kabashima-2003b}, 
the time evolution of $\pi_1^t(x)$ and $\pi_2^t(\x)$ can be written down in an iterative form as follows,
\begin{subequations}
\begin{align}
    \label{eq:1}&\pi_1^{t+1}(x)=\int\prod^{P-1}_{\mu=1}\dif{\x_\mu}\pi^t_2(\x_\mu)\delta\left(x-\tanh\left(\sum^{P-1}_{\mu=1}\x_\mu\right)\right),\\
    \label{eq:2}&\pi_2^{t}(\x)=\int \prod^{N-1}_{l=1}\dif x_l \pi_1^t(x_l)\left\langle\delta\left(\x-\frac{\xi^\mu}{\sqrt{N\sigma_\mu}}\frac{G\left(X_\mu\right)}{H\left(X_\mu\right)}\right)
    \right\rangle_{\bm{\xi}},\\
    &X_\mu\equiv -\frac{\sum_{l=1}^{N-1}\xi_l^\mu x_l/\sqrt{N}}{\sqrt{1-\sum_{l=1}^{N-1}x_l^2/N}}=-\frac{w_\mu}{\sqrt{\sigma_\mu}},
\end{align}
\end{subequations}
where $\left\langle\cdots\right\rangle$ represents the disorder average over $\bm{\xi}$. 
Note that $\xi^\mu$ is independent of the pattern entries in the sum of $X_\mu$.

We then introduce an auxiliary field $h_{l\to\mu}^t=\sum_{\nu\neq \mu}u_{\nu\to l}^t=\tanh^{-1}(m_{l\to\mu}^t)$ and its macroscopic distribution $\rho(h)$. More precisely,
\begin{equation}
  \rho^t(h)=\frac{1}{NP}\sum^N_{l=1}\sum^P_{\mu=1}\delta\left(h-h_{l\to\mu}^t\right).
\end{equation}
When $P$ becomes infinite (e.g., $P\propto N$), due to the central limit theorem, the distribution of the auxiliary field can be regarded as a Gaussian distribution:
\begin{equation}
    \label{eq:3}\rho^t(h)=\int\prod^{P-1}_{\mu=1}\dif \x_\mu \pi_2^t(\x_\mu)\delta\left(h-\sum^{P-1}_{\mu=1}\x_\mu\right)\approx
    \frac{1}{\sqrt{2\pi F^t}}\exp{\left[-\frac{\left(h-E^t\right)^2}{2F^t}\right]},
\end{equation}
where $E^t$ and $F^t$ are the mean and variance of the Gaussian distribution $\rho(h)$, respectively.
In fact, $E^t=0$ because of the setting that $\xi^\mu$ takes $\pm 1$ with equal probabilities.
With the expression of $\rho^t(h)$, we get $\pi_1^{t+1}(x)=\int\dif h \rho^t(h)\delta\left(x-\tanh(h)\right)$ for \reff{eq:1}.
Plugging this expression into \reff{eq:2} and using \reff{eq:3}, we obtain a compact expression for the update of $F^t$ as
\begin{subequations}\label{cav}
\begin{align}
    F^{t+1}=&\frac{\alpha}{1-Q^t}\int\Dif z \left(\frac{G\left(-\sqrt{\frac{Q^t}{1-Q^t}}z\right)}{H\left(-\sqrt{\frac{Q^t}{1-Q^t}}z\right)}\right)^2,\\
    Q^t=&\int\Dif z\tanh^2(\sqrt{F^t}z).
\end{align}
\end{subequations}
We leave the technical details of this derivation to Appendix~\ref{app-a}.
Note that this result is exactly identical to the saddle point equation under the replica symmetric assumption, which we shall briefly introduce in Sec.~\ref{replica} .

\section{Microscopic instability of the algorithmic iteration}
In this section, we turn to the analysis of the microscopic stability of the BP equations at a fixed point. 
Provided that a field fluctuation $\delta h^t_{l\to\nu}$ is introduced around the fixed point $m^t_{l\to\nu}=m_{l\to\nu}$,
the time evolution of $\delta h^t_{l\to\nu}$ is computed as 
\begin{equation}
    \label{eq:4}\delta h_{l\to\nu}^t = \sum_{\mu\neq\nu}\delta u_{\mu\to l} = \sum_{\mu\neq\nu} \frac{\xi^\mu_l}{\sqrt{N}} \left[L\delta w_{\mu\to l}+K \delta\sigma_{\mu\to l} \right],
\end{equation}
where
\begin{subequations}
\label{eq:4sp}
\begin{align}
    &\delta w_{\mu\to l}\equiv\frac{1}{\sqrt{N}}\sum_{i\neq l}\xi^\mu_i(1-m_{i\to\mu}^2)\delta h_{i\to\mu},\\
    &\delta \sigma_{\mu\to l}\equiv-\frac{2}{N}\sum_{i\neq l}m_{i\to\mu}\left(1-m_{i\to\mu}^2\right)\delta h_{i\to\mu},\\
    &K\equiv\left(\frac{w_{\mu\to l}^2}{\sigma_{\mu\to l}}+\frac{w_{\mu\to l}}{\sqrt{\sigma_{\mu\to l}}}\frac{G(X_\mu)}{H(X_\mu)}-1\right)\frac{G(X_\mu)}{H(X_\mu)}\frac{1}{2\sigma_{\mu\to l}^{\frac32}},\\
    &L\equiv-\frac{1}{\sigma_{\mu\to l}} \left(\frac{w_{\mu\to l}}{\sqrt{\sigma_{\mu\to l}}}\frac{G(X_\mu)}{H(X_\mu)}+\frac{G^2(X_\mu)}{H^2(X_\mu)}\right).
\end{align}
\end{subequations}
Note that in the right hand side of Eq.~(\ref{eq:4sp}), all messages or perturbations refer to their values at a previous step ($t-1$).
In the following analysis (including Appendix~\ref{app-a}), we omit this time index.
We then define 
the macroscopic distribution of $\delta h^t_{l\to\nu}$ as $f^t(y)$ \cite{kabashima-2003b}.
Due to the central limit theorem, $f^t(y)$ can be assumed to be a Gaussian form, i.e.,
\begin{equation}
    \label{eq:5}f^t(y)=\frac{1}{NP}\sum^{N}_{l=1}\sum^{P}_{\mu=1}\delta\left(y-\delta h_{l\to\mu}^t\right)\approx \frac{1}{\sqrt{2\pi b^t}}\exp{\left[-\frac{\left(y-a^t\right)^2}{2b^t}\right]},
\end{equation}
where $a^t$ and $b^t$ are the mean and variance of the distribution, respectively.
The time evolution of $f^t(y)$ is provided by a functional equation as follows,
\begin{equation}
    \label{eq:6}f^{t+1}(y)=\int\prod^P_{\mu=1}\prod^N_{l=1}\dif y_{l\to\mu} f^t(y_{l\to\mu})\left\langle\delta\left(y-\sum^{P-1}_{\mu=1} \frac{\xi^\mu}{\sqrt{N}} \left[L\delta w_\mu+K \delta\sigma_\mu \right] \right)\right\rangle_{\{x_{l\to\mu}\},\bm{\xi}}.
\end{equation}

Following the similar spirit as before, $a^t$ is zero. We thus only need to focus on the update of $b^t$. We provide details of derivation of this update in Appendix~\ref{app-a}.
We finally obtain
\begin{equation}
    \label{eq:7}b^{t+1}=\frac{\alpha}{\left(1-Q^t\right)^2}\int\Dif z\left(\frac{G(Z)}{H(Z)} \right)^2\left(Z-\frac{G(Z)}{H(Z)}\right)^2\int\Dif z\frac{1}{\cosh^4(\sqrt{F^t}z)}b^t\equiv\gamma b^t,
\end{equation}
where $Z\equiv\sqrt{Q^t/(1-Q^t)}z$.
When $\gamma<1$, $b^t$ converges to zero after iteration, indicating that the initially-introduced fluctuation of the 
auxiliary field will eventually vanish.
On the contrary, when $\gamma>1$, $b^t$ would grow with iteration, which implies that 
the fluctuation will be amplified, leading to the instability of the fixed point. 
Therefore, \reff{eq:7} provides the critical condition of the instability with respect to the growth of $b^t$, i.e.,
\begin{equation}
    \label{eq:21}\frac{\alpha}{\left(1-Q\right)^2}\int\Dif z\left(\frac{G(Z)}{H(Z)} \right)^2\left(Z-\frac{G(Z)}{H(Z)}\right)^2\int\Dif z\frac{1}{\cosh^4(\sqrt{F}z)} = 1.
\end{equation}

\section{Equilibrium properties via replica trick}
\label{replica}
In this section, we apply the replica trick to analyze the equilibrium properties of the binary perceptron. 
In the thermodynamic limit, the free energy has the self-averaging property, i.e., the distribution of the free energy for different realizations of learning is peaked at the typical value.
Thus, we can compute the disorder-average given by
$-\beta f=\left\langle\ln \mathrm{Z}\right\rangle$, where the average is carried out with respect to i.i.d random patterns.
In fact, this disorder-average is very hard to compute. However,
by introducing $n$ replicas of the original learning system and then setting $n\to0$, we can obtain the free energy of the system in a mathematically concise way~\cite{HH-2022}:
\begin{equation}
    \label{eq:8}-\beta f=\lim_{n\to 0,N\to\infty}\frac{\ln{\left\langle \mathrm{Z}^n\right\rangle}}{nN}=\lim_{n\to0}\frac{\ln e^{NF_{\max}}}{nN}=\lim_{n\to0}\frac{F_{\max}}{n}.
\end{equation}

Here, we are interested in the zero-temperature limit (focusing on ground states). Therefore, Equation~(\ref{eq:8}) is actually the entropy counting the number of solutions to the perceptron learning.
By introducing replicas (copies of the system), we transfer a direct intractable
treating of complex interactions in learning to handling the overlap matrix of states, which can be tackled by physics approximations, e.g., the RS ansatz in which 
the overlap does not depend on specific replica index  (permutation symmetry).  An intuitive picture is that the RS ansatz is consistent with the delta-like distribution of messages on
each link of the factor graph, and the broadening of the distribution (under the message perturbation)
leads to the mathematical instability of the saddle point. We will come back to this point at the end of Sec.~\ref{secrsb}.

To compute $\ln{\left\langle \mathrm{Z}^n\right\rangle}$, we introduce $n$ replicated synaptic weight vectors
$\bb{J}^a(a=1,\dots,n)$ as follows
\begin{equation}
    \begin{split}
        \label{eq:9}\left\langle \mathrm{Z}^n \right\rangle &= \left\langle \sum_{\{\bb{J}^a\}}\prod_{a,\mu}\Theta\left(\frac{1}{\sqrt{N}}\sum_i J^a_i \xi^\mu_i\right) \right\rangle\\
        &=\int\prod_{a<b}\frac{\dif q^{ab} \dif\hat{q}^{ab}}{2\pi \mi/N }\exp\left[-N\sum_{a<b}q^{ab}\hat{q}^{ab}+N\alpha G_0(\{q^{ab}\})+NG_1(\{\hat{q}^{ab}\})\right],
    \end{split}
\end{equation}
where we have introduced the state overlap $q^{ab}=\frac{1}{N}\sum_i J_i^a J_i^b$ and its associated conjugated counterpart $\hat{q}^{ab}$. The expressions of $G_0(\{q^{ab}\})$ (energy term)
and $G_1(\{\hat{q}^{ab}\})$ (entropy term) are given as follows~\cite{Krauth-1989,Huang-JPA2013}
\begin{subequations}
\begin{align}
    \label{eq:10}G_0(\{q^{ab}\})&=\ln{\int\prod_a\frac{\dif\lambda^a}{2\pi}}\int_0^\infty\dif t^a e^{\mi\sum_a\lambda^a t^a-\sum_{a<b}q^{ab}\lambda^a\lambda^b-\frac{1}{2}\sum_a(\lambda^a)^2},\\
    G_1(\{\hat{q}^{ab}\})&=\ln\sum_{\{J^a\}}e^{\sum_{a<b}\hat{q}^{ab}J^a J^b}.
\end{align}
\end{subequations}
Plugging \reff{eq:9} into \reff{eq:8}, we get the entropy
\begin{equation}
    \label{eq:11}s=\lim_{n\to 0}\frac{1}{n}\max\left[-\sum_{a<b}q^{ab}\hat{q}^{ab}+\alpha G_0(\{q^{ab}\})+G_1(\{\hat{q}^{ab}\})\right].
\end{equation}
Under the RS ansatz $q^{ab}=q, \hat{q}^{ab}=\hat{q}$ for $a\neq b$ , the extremization of \reff{eq:11} gives rise to the following saddle-point equations:
\begin{subequations}
\begin{align}
    q=&\int\Dif z\tanh^2(\sqrt{\hat{q}}z),\\
    \hat{q}=&\frac{\alpha}{1-q}\int\Dif z \left(\frac{G\left(-\sqrt{\frac{q}{1-q}}z\right)}{H\left(-\sqrt{\frac{q}{1-q}}z\right)}\right)^2.
\end{align}
\end{subequations}
These saddle point equations are again identical to Eq.~(\ref{cav}) derived from the BP equation.

\section{Instability of the replica symmetric solution}
\label{secrsb}
The stability of the RS solution requires that the eigenvalues of the Hessian matrix (the second derivative matrix) of $F_{\max}$ must be negative.
The sign of the eigenvalues of this matrix evaluated at the RS solution tells us all the information about the stability~\cite{AT-1978}. We first introduce
 $\eta^{ab}$ and $\epsilon^{ab}$ as the fluctuations around the RS solution as
 \begin{subequations}
\begin{align}
    q^{ab}=q+\eta^{ab},\\
    \hat{q}^{ab}=\hat{q}+\epsilon^{ab}.
\end{align}
\end{subequations}
By taking the Taylor expansion, we obtain $\frac{1}{2}\Delta$ as the second order terms of $F_{\max}$, where
\begin{equation}
    \label{eq:12}\Delta\equiv\left.\alpha\sum_{\alpha\beta,\gamma\delta}\frac{\partial^2G_0}{\partial q^{\alpha\beta}\partial q^{\gamma\delta}}\right|_{\eta^{\alpha\beta},\eta^{\gamma\delta}=0}\eta^{\alpha\beta}\eta^{\gamma\delta}-
    \left.\sum_{\alpha\beta,\gamma\delta}\eta^{\alpha\beta}\epsilon^{\gamma\delta}+\sum_{\alpha\beta,\gamma\delta}\frac{\partial^2G_1}{\partial \hat{q}^{\alpha\beta}\partial \hat{q}^{\gamma\delta}}\right|_{\epsilon^{\alpha\beta},\epsilon^{\gamma\delta}=0}\epsilon^{\alpha\beta}\epsilon^{\gamma\delta},
\end{equation}
where the prefactor $\alpha$ in the first term is the loading rate, the superscript of the order parameters indicates the replica index, $G_0=G_0(\{q^{\alpha\beta}\})$, and $G_1=G_1(\{\hat{q}^{\alpha\beta}\})$. In other words, the Hessian matrix looks like 
\begin{equation}
    \begin{bmatrix}
        \alpha \mathbf{H}_0 & -\mathbf{I}\\-\mathbf{I} &\mathbf{H}_1
    \end{bmatrix}
\end{equation}
composed of four $\frac{n(n-1)}{2}\times\frac{n(n-1)}{2}$ blocks. $H_0^{(\alpha\beta)(\gamma\delta)}=\frac{\partial G_0}{\partial q^{\alpha\beta}\partial q^{\gamma\delta}}$,
$H_1^{(\alpha\beta)(\gamma\delta)}=\frac{\partial G_1}{\partial \hat{q}^{\alpha\beta}\partial \hat{q}^{\gamma\delta}}$, and $\mathbf{I}$ is an identity matrix.

Following the Gardner's analysis~\cite{Gardner-1988}, we first consider the problem of diagonalizing the matrices of $\mathbf{H}_0$ and $\mathbf{H}_1$ separately.
We then use the symmetry structure with respect to permutation of replica indices. The associated eigenvectors can be divided into three types (see details in
Appendix~\ref{app-b} and Appendix~\ref{app-c}). The first type are symmetric for all indices. 
The second type are symmetric for all but one specific index, and the third type are symmetric for all but two specific indices. 
In the limit of $n\to0$, the second type of eigenvectors coincides with the first type of eigenvectors.
The first type of eigenvectors defines the longitudinal fluctuations within the RS subspace~\cite{Gardner-1988,Engel-2001}.
This stability is already guaranteed by optimizing the action $F_{\max}$. In other words, the sufficient condition for $\lambda_{1,2}<0$ is equivalent to the saddle point equation~\cite{Engel-2001}.

Therefore, only the third type of eigenvectors leads to the instability of the RS solution. This type of eigenvectors corresponds to the instability that is able to take the stationary point outside the RS subspace, capturing the transverse fluctuations.
Supposed that the eigenvalues of these eigenvectors for $\mathbf{H}_0$ and $\mathbf{H}_1$ are $\gamma_1$ and $\gamma_2$, respectively.
The related eigenvalues that cause the instability of the RS solution are given by the two eigenvalues of the following matrix
\begin{equation}
    \begin{bmatrix}
        \alpha\gamma_1 & -1\\-1 &\gamma_2
    \end{bmatrix}.
\end{equation}
The sign of the determinant determines the stability of the RS solution, i.e., the RS solution is stable only when $\alpha\gamma_1\gamma_2<1$.
When $\alpha\to0$, the determinant of this matrix is given by $\alpha\gamma_1\gamma_2-1=-1$, which means that the product of eigenvalues is negative.
Therefore, in this limit, the RS solution is correct as expected.
When $\alpha$ increases above a critical value, the sign of the determinant changes, which means that one of these eigenvalues changes its sign, thereby breaking
the stability of the RS solution.

\begin{figure}
	\centering
	\includegraphics[bb=58 45 669 530,scale=0.5]{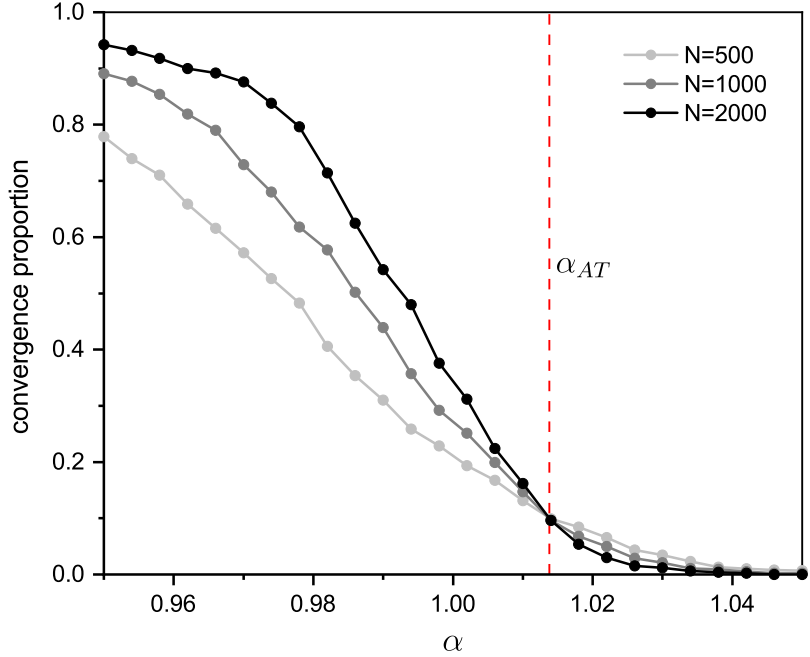}
	\caption{The convergence proportion of the BP algorithm versus loading rate. 
	The red dash line marks the $\alpha_{AT}$ computed by the stability condition equation of the RS solution. 
	The three curves for different network sizes intersect at a point that coincides with $\alpha_{AT}$. For each
	data point on the curves, we simulate $M$ instances of binary perceptron. $M=2\,000$ for $N=500$, $1\,000$ for $N=1000$, and $500$ for $N=2\,000$.
	}
	\label{comp}
\end{figure}

According to the calculation details in the Appendix~\ref{app-b}, we have
\begin{subequations}
\begin{align}
    \gamma_1&=\frac{1}{\left(1-q\right)^2}\int\Dif z\left(\frac{G(Z)}{H(Z)}\right)^2\left(Z-\frac{G(Z)}{H(Z)}\right)^2,\\
    \gamma_2&=\int\Dif z\frac{1}{\cosh^4\left(\sqrt{\hat{q}}z\right)},
\end{align}
\end{subequations}
where $Z=\sqrt{q/(1-q)}z$.
Therefore, the critical condition for the transition to replica symmetry breaking is specified by
\begin{equation}
    \label{eq:22}\alpha\gamma_1\gamma_2=\frac{\alpha}{\left(1-q\right)^2}\int\Dif z\left(\frac{G(Z)}{H(Z)} \right)^2\left(Z-\frac{G(Z)}{H(Z)}\right)^2\int\Dif z\frac{1}{\cosh^4(\sqrt{\hat{q}}z)}=1.
\end{equation}
We thus conclude that \reff{eq:22} is identical to \reff{eq:21}, which suggests
that the equivalence between algorithmic instability and transition to replica symmetry breaking can be established
in perceptron learning systems. The replica symmetry breaking captures a hierarchical organization of replicas. In physics, this actually corresponds to
the decomposition of the Gibbs measure into (exponentially or sub-exponentially) many pure states~\cite{MM-2009}.

We finally carry out a numerical simulation to check whether the theoretical instability coincides that obtained by running BP in specific instances.
As shown in Fig.~\ref{comp}, we observe the theoretical prediction, namely the Almeida-Thouless (AT)~\cite{AT-1978} loading rate ($\alpha_{\rm AT}$) matches well the numerical
estimation. The theoretical prediction is computed by solving Eq.~(\ref{cav}) and Eq.~(\ref{eq:21}). During simulations, we estimate the convergence proportion as
the fraction of instances for which the BP iteration converges within a prescribed criterion (e.g., all updated messages within a small deviation from the values at the previous
iteration). It is expected from the plot that in the thermodynamic limit, the BP iteration does not converge beyond $\alpha_{\rm AT}$ with the probability tending to one.

\section{Conclusion}
In this work, from a physics perspective, we prove that the stability of the learning algorithm, derived using physically intuitive cavity method, is connected to the stability of the replica symmetric saddle point solution of 
the model. The equivalence between physically intuitive cavity method and the mathematically concise replica method was also explored in spin interaction systems~\cite{kabashima-2003b},
information transmission
systems~\cite{kabashima-2003a}, linear estimation problems such as compressed sensing~\cite{Donoho-2009,Krzakala-2012,Lenka-2016}, and
spectra estimation of
random sparse matrices~\cite{Franti-2011}. Our proof adds another evidence of this equivalence in perceptron learning systems, by claiming rigorously (in the 
thermodynamic limit) the one-to-one correspondence between the BP instability and the AT instability of the equilibrium saddle point.

Our framework shows that the cumbersome replica analysis could be avoided in studying learning systems, e.g., stability analysis considered in this work. Therefore,
this work would hopefully inspire further studies on landscape analysis~\cite{Huang-2014pre,Baldassi-2015}, unsupervised learning~\cite{Hou-2019,Huang-2020}, and even 
deep learning, e.g., a current hot topic of learning in overparameterized neural
networks~\cite{baldassi-2021}.

\begin{acknowledgments}
We would like to thank other PMI members for discussions.
This research was supported by the National Natural Science Foundation of China for
Grant numbers 12122515 and 11805284 (HH). 
\end{acknowledgments}

\appendix
\section{Instability analysis of the BP iteration}
\label{app-a}
The iterative equation for the field distribution reads
\begin{subequations}
\begin{align}
    \rho^{t+1}(h)&=\int\prod^{P-1}_{\mu=1}\prod^{N-1}_{l=1}\dif{h_{l\to\mu}}\rho^t(h_{l\to\mu})\left\langle \delta\left( h-\sum_{\mu=1}^{P-1}\frac{\xi^\mu}{\sqrt{N\sigma_\mu}}\frac{G\left(X_\mu\right)}{H\left(X_\mu\right)}\right)\right\rangle_{\bm{\xi}},\\
    X_\mu&= -\frac{\sum_{l=1}^{N-1}\xi_l^\mu \tanh(\h)/\sqrt{N}}{\sqrt{1-\sum_{l=1}^{N-1}\tanh^2(\h)/N}}=-\frac{w_\mu}{\sqrt{\sigma_\mu}}.
\end{align}
\end{subequations}
Notice that $\xi^\mu$ is independent of $\xi^\mu_l$. We can thus calculate the average with respect to $\xi^\mu$ and $\xi^\mu_l$ separately.
Due to the zero mean of $\xi^\mu$, $E^t=\int h\rho^t(h)\dif h=0$.
Because $\xi_l^\mu=\pm 1$, we introduce a transformation $\h\to \xi_l^\mu \h$, which gives rise to
\begin{equation}
    X_\mu= -\frac{\sum_{l=1}^{N-1} \tanh(\h)/\sqrt{N}}{\sqrt{1-\sum_{l=1}^{N-1}\tanh^2(\h)/N}}.
\end{equation}
Then the variance reads,
\begin{equation}
\begin{split}
    \label{eq:14}F^{t+1}&=\int h^2 \rho^{t+1}(h)\dif h =\int\prod^{P-1}_{\mu=1}\prod^{N-1}_{l=1}\dif{\h}\rho^t(\h) \sum_{\mu=1}^{P-1}\left[\frac{1}{\sqrt{N\sigma_\mu}}\frac{G\left(X_\mu\right)}{H\left(X_\mu\right)}\right]^2\\
    &=\alpha\int\prod^{N-1}_{l=1}\dif{h_{l}}\rho^t(h_{l}) \left[\frac{1}{\sqrt{\sigma}}\frac{G\left(X\right)}{H\left(X\right)}\right]^2\\
    &=\alpha\mathbb{E}\left[\left(\frac{1}{\sqrt{\sigma}}\frac{G\left(X\right)}{H\left(X\right)}\right)^2\right],
\end{split}
\end{equation}
where we have used the i.i.d property of the random patterns.

When $N\to\infty$, due to the law of large numbers, we have
\begin{equation}\label{app1}
    \lim_{N\to\infty}\frac{1}{N}\sum^{N-1}_{l=1}\tanh^2(h_l)=\mathbb{E}\left[\tanh^2(h_l)\right]=\int\Dif z\tanh^2(\sqrt{F^t}z)\equiv Q^t.
\end{equation}
Due to the central limit theorem, we also have
\begin{equation}\label{app2}
    \begin{split}
    \lim_{N\to\infty}\frac{1}{\sqrt{N}}\sum^{N-1}_{l=1}\tanh(h_l) &= \sqrt{\mathbb{E}\left[\tanh^2(h_l)\right]}z+\mathbb{E}\left[\tanh(h_l)\right]\\
    &=\sqrt{\mathbb{E}\left[\tanh^2(h_l)\right]}z=\sqrt{Q^t}z,
\end{split}
\end{equation}
where $z\sim \mathcal{N}\left(0,1\right)$. Therefore,
\begin{equation}\label{app3}
    \begin{split}
        \lim_{N\to\infty}X_\mu=-\frac{\lim_{N\to\infty}\sum_{l=1}^{N-1}\tanh(\h)/\sqrt{N}}{\sqrt{1-\lim_{N\to\infty}\sum_{l=1}^{N-1}\tanh^2(\h)/N}}=-\sqrt{\frac{Q^t}{1-Q^t}}z\equiv-Z.
    \end{split}
\end{equation}
Plugging Eqs.~(\ref{app1})-(\ref{app3}) into \reff{eq:14}, we have
\begin{equation}
    F^{t+1}=\frac{\alpha}{1-Q^t}\int\Dif z \left(\frac{G\left(-\sqrt{\frac{Q^t}{1-Q^t}}z\right)}{H\left(-\sqrt{\frac{Q^t}{1-Q^t}}z\right)}\right)^2.
\end{equation}

Next, we calculate the time evolution of $a^t$ and $b^t$. Because of the zero-mean of $\xi^\mu$, it can also be proved that $a^t=0$. 
In addition, $b^{t+1}$ is the second-order moment of $f(y)$, i.e., 
\begin{equation}
\begin{split}
    b^{t+1}&=\int y^2f^{t+1}(y)\dif y\\
    &=\int\prod^P_{\mu=1}\prod^N_{l=1}\dif \y f^t(\y)\left\langle \left(\sum^{P-1}_{\mu=1} \frac{\xi^\mu}{\sqrt{N}} \left[L\delta w_\mu+K \delta\sigma_\mu \right] \right)^2\right\rangle_{\{\xl\},\bm{\xi}}\\
    &=\frac{1}{N}\int\prod^P_{\mu=1}\prod^N_{l=1}\dif \y f^t(\y)\sum^{P-1}_{\mu=1} \left\langle \left[L\delta w_\mu+K \delta\sigma_\mu \right]^2\right\rangle_{\{\xl\},\bm{\xi}}\\
    &=\frac{1}{N}\int\prod^P_{\mu=1}\prod^N_{l=1}\dif \y f^t(\y)\sum^{P-1}_{\mu=1} \left\langle W_\mu\right\rangle_{\{\xl\}},
\end{split}
\end{equation}
where
\begin{equation}
    W_\mu\equiv\left\langle\left[\frac{L}{\sqrt{N}}\sum_{l=1}^{N-1}\xi^\mu_l(1-\xl^2)\y-\frac{2K}{N}\sum_{l=1}^{N-1}\xl\left(1-\xl^2\right)\y\right]^2\right\rangle_{\bm{\xi}}.
\end{equation}
Performing the distribution-preserved transformation $\xl\to\xi^\mu_l\xl$ and neglecting the higher-order small terms in the large-N limit,
we arrive at
\begin{equation}
    \begin{split}
        W_\mu&=\left\langle\left[\frac{L}{\sqrt{N}}\sum_{l=1}^{N-1}\xi^\mu_l(1-\xl^2)\y-\frac{2K}{N}\sum_{l=1}^{N-1}\xi^\mu_l\xl\left(1-\xl^2\right)\y\right]^2\right\rangle_{\bm{\xi}}\\
        &\simeq\sum_{l=1}^{N-1}\left[\frac{L}{\sqrt{N}}\right]^2(1-\xl^2)^2\y^2,
    \end{split}
\end{equation}
and immediately we get
\begin{equation}
    \begin{split}
        b^{t+1}&=\lim_{N\to\infty}b^t\alpha\left\langle\sum_{l=1}^{N-1}\left[\frac{L}{\sqrt{N}}\right]^2(1-x_{l}^2)^2\right\rangle_{\{x_{l}\}}.
    \end{split}
\end{equation}

Note that
\begin{equation}
    \begin{split}
        L&=-\lim_{N\to\infty}\frac{1}{\sigma}\left(\frac{w}{\sqrt{\sigma}}\frac{G(X)}{H(X)}+\frac{G^2(X)}{H^2(X)}\right)\\
        &=-\frac{1}{\left(1-Q^t\right)} \left(\frac{G(-Z)}{H(-Z)}Z+\frac{G^2(-Z)}{H^2(-Z)}\right),
    \end{split}
\end{equation}
and
\begin{equation}
    \begin{split}
        \lim_{N\to\infty}\frac{1}{N}\sum_{l=1}^{N-1}(1-x_{l}^2)^2 &= 1-2\mathbb{E}\left[x_l^2\right]+\mathbb{E}\left[x_l^4\right]\\
        &=1-2\int\Dif z\tanh^2(\sqrt{F^t}z)+\int\Dif z\tanh^4(\sqrt{F^t}z)\\
        &=\int\Dif z\frac{1}{\cosh^4(\sqrt{F^t}z)}.
    \end{split}
\end{equation}
Finally, we get
\begin{equation}
    \begin{split}
        b^{t+1}=\frac{\alpha}{\left(1-Q^t\right)^2}\int\Dif z\left(\frac{G(Z)}{H(Z)} \right)^2\left(Z-\frac{G(Z)}{H(Z)}\right)^2\int\Dif z\frac{1}{\cosh^4(\sqrt{F^t}z)} b^t,
    \end{split}
\end{equation}
where a statistically invariant change $z\to-z$ has been made.

\section{Derivation of saddle point equations}
\label{app-sde}
In this section, we show explicitly how the replica computation is carried out.
Applying the RS ansatz $q^{ab}=q,\hat{q}^{ab}=\hat{q}$ for $a\neq b$ to the energy term, we obtain
\begin{equation}
    \begin{split}
        G_0&(q)=\ln{\int\prod_a\frac{\dif\lambda^a}{2\pi}}\int_0^\infty\dif t^a e^{\mi\sum_a\lambda^a t^a-\frac{1}{2}q(\sum_{a}\lambda^a)^2-\frac{1}{2}(1-q)\sum_a(\lambda^a)^2}\\
        &=\ln{\int\Dif z\int\prod_a\frac{\dif\lambda^a}{2\pi}}\int_0^\infty\dif t^a e^{\mi\sum_a\lambda^a t^a-\mi\sum_{a}\lambda^a\sqrt{q}z-\frac{1}{2}(1-q)\sum_a(\lambda^a)^2}\\ 
        &=\ln{\int\Dif z\int\prod_a\frac{\dif\lambda^a}{2\pi}}\int_{-\sqrt{q}z}^\infty\dif t^a e^{\mi\sum_a\lambda^a t^a-\frac{1}{2}(1-q)\sum_a(\lambda^a)^2}\\
        \label{eq:25}&=\ln{\int\Dif z \left[ \int\frac{\dif\lambda}{2\pi}\int_{-\sqrt{q}z}^\infty\dif t e^{\mi\lambda t-\frac{1}{2}(1-q)\lambda^2}\right]^n}=\ln{\int\Dif z \left[ \int \frac{\dif\lambda}{2\pi}\int_{-\sqrt{\frac{q}{1-q}}z}^\infty\dif t e^{\mi\lambda t-\frac{1}{2}\lambda^2}\right]^n}\\
        &=\ln{\int\Dif z \left[ \int^\infty_{-\sqrt{\frac{q}{1-q}}z}\Dif t\right]^n}=\ln{\int\Dif z \left[ H\left(-\sqrt{\frac{q}{1-q}}z\right)\right]^n},
    \end{split}
\end{equation}
where we have rescaled $t=t\sqrt{1-q}$ and $\lambda=\lambda/\sqrt{1-q}$.
Then we compute the entropy term $G_1(\hat{q})$ as
\begin{equation}
    \begin{split}
    G_1(\hat{q})&=\ln\sum_{\{J^a\}}e^{\sum_{a<b}\hat{q}^{ab}J^a J^b}=\ln\sum_{\{J^a\}}e^{\hat{q}\sum_{a<b}J^a J^b}=\ln\sum_{\{J^a\}}e^{\frac{\hat{q}}{2}(\sum_{a}J^a)^2 -\frac{\hat{q}n}{2}}\\
    &=\ln\int\Dif z\sum_{\{J^a\}} e^{\sqrt{\hat{q}}z \sum_{a}J^a -\frac{\hat{q}n}{2}}=\ln\int\Dif z e^{-\frac{\hat{q}n}{2}}\sum_{\{J^a\}}\prod_a e^{\sqrt{\hat{q}}zJ^a}\\
    &=-\frac{\hat{q}n}{2}+\ln\int\Dif z\prod_a\left[\sum_{J^a}e^{\sqrt{\hat{q}}zJ^a}\right]=-\frac{\hat{q}n}{2}+\ln\int\Dif z\left[2\cosh{\sqrt{\hat{q}}z}\right]^n.
    \end{split}
\end{equation}
Therefore, the entropy of the model turns out to be
\begin{equation}
    \begin{split}
    s&=\lim_{n\to 0}\frac{1}{n}\max\left[ -\frac{n(n-1)}{2}q\hat{q}-\frac{n}{2}\hat{q}+\ln{\int\Dif z \left[ H\left(-\sqrt{\frac{q}{1-q}}z\right)\right]^n}+\ln\int\Dif z\left[2\cosh{\sqrt{\hat{q}}z}\right]^n\right]\\
    &=\frac{q\hat{q}}{2}-\frac{\hat{q}}{2}+\int\Dif z \ln{\left[ H\left(-\sqrt{\frac{q}{1-q}}z\right)\right]}+\int\Dif z\ln\left[2\cosh{\sqrt{\hat{q}}z}\right].
\end{split}
\end{equation}
Finally, we arrive at the saddle point equations as follows
\begin{align}
    &\frac{\partial s}{\partial\hat{q}}=0\quad\Rightarrow\quad q=\int\Dif{z}\tanh^2(\sqrt{\hat{q}}z),\\
    &\frac{\partial s}{\partial q}=0 \quad\Rightarrow\quad \hat{q}=\frac{\alpha}{1-q}\int\Dif t \left(\frac{G\left(-\sqrt{\frac{q}{1-q}}t\right)}{H\left(-\sqrt{\frac{q}{1-q}}t\right)}\right)^2.
\end{align}

\section{Instability analysis of the RS solution}
\label{app-b}
Considering the perturbation on the order parameters, we write the energy 
term $G_0$ as
\begin{equation}
\begin{split}
    G_0&=\ln{\int\prod_a\frac{\dif\lambda^a}{2\pi}}\int_0^\infty\dif t^a e^{\mi\sum_a\lambda^a t^a-\frac{1}{2}q(\sum_{a}\lambda^a)^2-\frac{1}{2}(1-q)\sum_a(\lambda^a)^2-\sum_{a<b}\eta^{ab}\lambda^a\lambda^b}\\
    &=\ln{\int\Dif z\int\prod_a\frac{\dif\lambda^a}{2\pi}}\int_0^\infty\dif t^a e^{\mi\sum_a\lambda^a t^a-\mi\sum_{a}\lambda^a\sqrt{q}z-\frac{1}{2}(1-q)\sum_a(\lambda^a)^2-\sum_{a<b}\eta^{ab}\lambda^a\lambda^b}\\
    &=\ln{\int\Dif z\int\prod_a\frac{\dif\lambda^a}{2\pi}}\int_{-\sqrt{q}z}^\infty\dif t^a e^{\mi\sum_a\lambda^a t^a-\frac{1}{2}(1-q)\sum_a(\lambda^a)^2-\sum_{a<b}\eta^{ab}\lambda^a\lambda^b},
\end{split}
\end{equation}
where we have shifted the integral variable $t^a\to t^a-\sqrt{q}z$. In addition,
we define $G_0'$ as
\begin{equation}
    G_0'\equiv\int\Dif z\ln{\int\prod_a\frac{\dif\lambda^a}{2\pi}}\int_{-\sqrt{q}z}^\infty\dif t^a e^{\mi\sum_a\lambda^a t^a-\frac{1}{2}(1-q)\sum_a(\lambda^a)^2-\sum_{a<b}\eta^{ab}\lambda^a\lambda^b}.
\end{equation}
When $n\to 0$ and $\eta^{ab}\to 0$, we have
\begin{equation}
    \begin{split}
        \lim_{\eta^{ab}\to0}\lim_{n\to 0}\frac{G_0}{G_0'}&=\lim_{n\to 0}\frac{\ln{\int\Dif z\int\prod_a\frac{\dif\lambda^a}{2\pi}}\int_{-\sqrt{q}z}^\infty\dif t^a e^{\mi\sum_a\lambda^a t^a-\frac{1}{2}(1-q)\sum_a(\lambda^a)^2}}{\int\Dif z\ln{\int\prod_a\frac{\dif\lambda^a}{2\pi}}\int_{-\sqrt{q}z}^\infty\dif t^a e^{\mi\sum_a\lambda^a t^a-\frac{1}{2}(1-q)\sum_a(\lambda^a)^2}}\\
        &=\lim_{n\to 0}\frac{\ln{\int\Dif z \left[ \int\frac{\dif\lambda}{2\pi}\int_{-\sqrt{q}z}^\infty\dif t e^{\mi\lambda t-\frac{1}{2}(1-q)\lambda^2}\right]^n}}{n\int\Dif z \ln{\left[ \int\frac{\dif\lambda}{2\pi}\int_{-\sqrt{q}z}^\infty\dif t e^{\mi\lambda t-\frac{1}{2}(1-q)\lambda^2}\right]}}=1.
    \end{split}
\end{equation}
Therefore, we can replace $G_0$ by $G_0'$ in the above two limits in \reff{eq:12}. We then get
\begin{equation}
    \label{eq:15}\left.H_0^{(\alpha\beta)(\gamma\delta)}\equiv\frac{\partial^2 G_0}{\partial q^{\alpha\beta}\partial q^{\gamma\delta}}\right|_{\eta^{\alpha\beta},\eta^{\gamma\delta}=0}=\left.\frac{\partial^2G_0'}{\partial \eta^{\alpha\beta}\partial \eta^{\gamma\delta}}\right|_{\eta^{\alpha\beta},\eta^{\gamma\delta}=0}=\left\langle\lambda^{\alpha}\lambda^{\beta}\lambda^{\gamma}\lambda^{\delta}\right\rangle-\left\langle\lambda^{\alpha}\lambda^{\beta}\right\rangle\left\langle\lambda^{\gamma}\lambda^{\delta}\right\rangle,
\end{equation}
where
\begin{equation}
    \left\langle f(\lambda)\right\rangle\equiv\int\Dif z\frac{\int\prod_a\frac{\dif\lambda^a}{2\pi}\int_{-\sqrt{q}z}^\infty\dif t^a f(\lambda)e^{\mi\sum_a\lambda^a t^a-\frac{1}{2}(1-q)\sum_a(\lambda^a)^2}}{\int\prod_a\frac{\dif\lambda^a}{2\pi}\int_{-\sqrt{q}z}^\infty\dif t^a e^{\mi\sum_a\lambda^a t^a-\frac{1}{2}(1-q)\sum_a(\lambda^a)^2}}.
\end{equation}

At the RS saddle point, \reff{eq:15} takes three possible values:
\begin{subequations}
\begin{align}
    P=H_0^{(\alpha\beta)(\alpha\beta)}&=\left\langle(\lambda^\alpha\lambda^\beta)^2\right\rangle-(\left\langle\lambda^\alpha\lambda^\beta\right\rangle)^2,\\
    Q=H_0^{(\alpha\beta)(\alpha\gamma)}&=\left\langle(\lambda^\alpha)^2\lambda^\beta\lambda^\gamma\right\rangle-\left\langle\lambda^{\alpha}\lambda^{\beta}\right\rangle\left\langle\lambda^{\alpha}\lambda^{\gamma}\right\rangle,\\
    R=H_0^{(\alpha\beta)(\gamma\delta)}&=\left\langle\lambda^{\alpha}\lambda^{\beta}\lambda^{\gamma}\lambda^{\delta}\right\rangle-\left\langle\lambda^{\alpha}\lambda^{\beta}\right\rangle\left\langle\lambda^{\gamma}\lambda^{\delta}\right\rangle,
\end{align}
\end{subequations}
where $\alpha,\beta,\gamma$ and $\delta$ are not equal with each other. 
We then compute the relevant moment terms as follows.
\begin{equation}
    \begin{split}
    \left\langle(\lambda^\alpha\lambda^\beta)^2\right\rangle&=\int\Dif z\frac{\int\prod_a\frac{\dif\lambda^a}{2\pi}\int_{-\sqrt{q}z}^\infty\dif t^a (\lambda^\alpha)^2(\lambda^\beta)^2 e^{\mi\sum_a\lambda^a t^a-\frac{1}{2}(1-q)\sum_a(\lambda^a)^2}}{\int\prod_a\frac{\dif\lambda^a}{2\pi}\int_{-\sqrt{q}z}^\infty\dif t^a e^{\mi\sum_a\lambda^a t^a-\frac{1}{2}(1-q)\sum_a(\lambda^a)^2}}\\
     &=\int\Dif z\frac{\int\frac{\dif\lambda^\alpha\dif\lambda^\beta}{\left(2\pi\right)^2}\int_{-\sqrt{q}z}^\infty\dif t^\alpha\dif t^\beta (\lambda^\alpha)^2(\lambda^\beta)^2 e^{\mi(\lambda^\alpha t^\alpha+\lambda^\beta t^\beta)-\frac{1}{2}(1-q)\left((\lambda^\alpha)^2+(\lambda^\beta)^2\right)}}{\int\frac{\dif\lambda^\alpha\dif\lambda^\beta}{\left(2\pi\right)^2}\int_{-\sqrt{q}z}^\infty\dif t^\alpha\dif t^\beta e^{\mi(\lambda^\alpha t^\alpha+\lambda^\beta t^\beta)-\frac{1}{2}(1-q)\left((\lambda^\alpha)^2+(\lambda^\beta)^2\right)}}\\
    &=\int\Dif z\frac{\int\frac{\dif\lambda^\alpha}{2\pi}\int_{-\sqrt{q}z}^\infty\dif t^\alpha (\lambda^\alpha)^2 e^{\mi(\lambda^\alpha t^\alpha)-\frac{1}{2}(1-q)(\lambda^\alpha)^2}}{\int\frac{\dif\lambda^\alpha}{2\pi}\int_{-\sqrt{q}z}^\infty\dif t^\alpha e^{\mi(\lambda^\alpha t^\alpha)-\frac{1}{2}(1-q)(\lambda^\alpha)^2}} \frac{\int\frac{\dif\lambda^\beta}{2\pi}\int_{-\sqrt{q}z}^\infty\dif t^\beta (\lambda^\beta)^2 e^{\mi(\lambda^\beta t^\beta)-\frac{1}{2}(1-q)(\lambda^\beta)^2}}{\int\frac{\dif\lambda^\beta}{2\pi}\int_{-\sqrt{q}z}^\infty\dif t^\beta e^{\mi(\lambda^\beta t^\beta)-\frac{1}{2}(1-q)(\lambda^\beta)^2}}\\
    &=\int\Dif z\left(\frac{\int\frac{\dif\lambda}{2\pi}\int_{-\sqrt{q}z}^\infty\dif t \lambda^2 e^{\mi\lambda t-\frac{1}{2}(1-q)\lambda^2}}{\int \frac{\dif\lambda}{2\pi}\int_{-\sqrt{q}z}^\infty\dif t e^{\mi\lambda t-\frac{1}{2}(1-q)\lambda^2}}\right)^2\\
    &=\int\Dif z \left(\overline{\lambda^2}\left[z\right]\right)^2,
    \end{split}
\end{equation}
where $\overline{f\left(\lambda\right)}[z]$ is defined as
\begin{equation}
    \overline{f(\lambda)}\left[z\right]=\frac{\int\frac{\dif\lambda}{2\pi}\int_{-\sqrt{q}z}^\infty\dif t f(\lambda) e^{\mi\lambda t-\frac{1}{2}(1-q)\lambda^2}}{\int \frac{\dif\lambda}{2\pi}\int_{-\sqrt{q}z}^\infty\dif t e^{\mi\lambda t-\frac{1}{2}(1-q)\lambda^2}}.
\end{equation}
Analogously, we obtain
\begin{subequations}
\begin{align}
    &\left\langle(\lambda^\alpha)^2\lambda^\beta\lambda^\gamma\right\rangle=\int\Dif z\left(\overline{\lambda}[z]\right)^2\overline{\lambda^2}[z],\\
    &\left\langle\lambda^{\alpha}\lambda^{\beta}\lambda^{\gamma}\lambda^{\delta}\right\rangle=\int\Dif{z}\left(\overline{\lambda}[z]\right)^4,
\end{align}
\end{subequations}
where $\overline{\lambda^2}\left[z\right]$ and $\overline{\lambda}\left[z\right]$ are computed as
\begin{subequations}
\begin{align}
    &\overline{\lambda}[z]=\frac{\mi}{\sqrt{1-q}}\frac{G(-Z)}{H(-Z)},\\
    &\overline{\lambda^2}[z]=\frac{1}{1-q}\frac{G\left(-Z\right)}{H\left(-Z\right)}Z,
\end{align}
\end{subequations}
where $Z=\sqrt{q/(1-q)}z$. 

Finally, we obtain $\gamma_1$ as
\begin{equation}
    \begin{split}
        \gamma_1&=P-2Q+R=\int\Dif z\left(\overline{\lambda^2}[z]-\left(\overline{\lambda}[z]\right)^2\right)^2\\
        &=\frac{1}{\left(1-q\right)^2}\int\Dif z\left(\frac{G(Z)}{H(Z)}\right)^2\left(Z-\frac{G(Z)}{H(Z)}\right)^2.
    \end{split}    
\end{equation}
To compute $\gamma_2$, we first define
\begin{subequations}
\begin{align}
    P'=&H_{1}^{(\alpha\beta)(\alpha\beta)}=1-\left(\left\langle J^\alpha J^\beta\right\rangle\right)^2,\\
    Q'=&H_1^{(\alpha\beta)(\alpha\gamma)}=\left\langle J^\beta J^\gamma\right\rangle-\left\langle J^\alpha J^\beta\right\rangle\left\langle J^\alpha J^\gamma \right\rangle,\\
    R'=&H_1^{(\alpha\beta)(\gamma\delta)}=\left\langle J^\alpha J^\beta J^\gamma J^\delta\right\rangle-\left\langle J^\alpha J^\beta\right\rangle\left\langle J^\gamma J^\delta \right\rangle,
\end{align}
\end{subequations}
where $\left\langle f(J)\right\rangle$ is defined as
\begin{equation}
    \left\langle f(J)\right\rangle=\int\Dif z \frac{\sum_{\{J^a\}}f(J) e^{\sqrt{\hat{q}}z \sum_{a}J^a}}{\sum_{\{J^a\}} e^{\sqrt{\hat{q}}z \sum_{a}J^a}}.
\end{equation}
We finally get
\begin{equation}
    \begin{split}
    \gamma_2&=P'-2Q'+R'=1-2\left\langle J^\beta J^\gamma\right\rangle+\left\langle J^\alpha J^\beta J^\gamma J^\delta\right\rangle\\
    &=1-2\int\Dif z\tanh^2\left(\sqrt{\hat{q}}z\right)+\int\Dif z\tanh^4\left(\sqrt{\hat{q}}z\right)\\
    &=\int\Dif z\frac{1}{\cosh^4\left(\sqrt{\hat{q}}z\right)}.
    \end{split}
\end{equation}

\section{Eigenvalues of the Hessian matrix}
\label{app-c}
Due to the symmetry with respect to permutation of replica indices,
there are three types of eigenvectors for the Hessian matrix $\bb{H}=\{H^{\left(\alpha\beta\right)\left(\gamma\delta\right)}\}$ \cite{AT-1978}. 
The first type of eigenvectors $\bm{\mu}_1$ has the following form
\begin{equation}
    \mu^{\alpha\beta} = a\quad\forall\quad\alpha<\beta.
\end{equation}
For all rows of $\bb{H}\bm{\mu}_1=\lambda_1\bm{\mu}_1$, the equations can be generally written as
\begin{equation}
    Pa+2(n-2)Qa+\frac{1}{2}(n-2)(n-3)Ra=\lambda_1a.
\end{equation}
While $n\to 0$, we obtain
\begin{equation}
    \lambda_1=P-4Q+3R.
\end{equation}

The second type of eigenvectors $\bm{\mu}_2$ has the following form
\begin{equation}
    \begin{split}
    &\mu^{\alpha\theta}=\mu^{\theta\beta}=b\qquad\alpha,\beta\neq\theta,\\
    &\mu^{\alpha\beta}=c\qquad\alpha,\beta\neq\theta,
    \end{split}
\end{equation}
where $\theta$ is the specific replica index.
From $\bb{H}\bm{\mu}_2=\lambda_2\bm{\mu}_2$, we obtain
\begin{equation}
    \label{eq:16}Pb+(n-2)Qb+(n-2)Qc+\frac{1}{2}(n-2)(n-3)Rc=\lambda_2 b.
\end{equation}
Because $\bb{H}$ is a symmetric matrix, the eigenvectors corresponding to different eigenvalues are orthogonal to each other. 
Therefore, $\bm{\mu}_1$ should be orthogonal to $\bm{\mu}_2$, leading to the following equation
\begin{equation}
    \label{eq:17}(n-1)ab+\frac{1}{2}(n-2)(n-1)ac=0.
\end{equation}
Using \reff{eq:16} and \reff{eq:17} and setting $n\to 0$, we get
\begin{equation}
    \lambda_2=P-4Q+3R.
\end{equation}
Due to the choice of one specific replica, this eigenvalue is ($n-1$)-fold degenerate.
We thus conclude that the eigenvalues of these two types of eigenvectors are the same in the limit of $n\to 0$. 
In fact, when $n\to 0$, $c$ will also converge to $b$, making the forms of the two types of eigenvectors the same.

The third type of eigenvectors $\bm{\mu}_3$ has the following form
\begin{equation}
    \begin{split}
    &\mu^{\theta\nu}=d,\\
    &\mu^{\alpha\nu}=\mu^{\nu\beta}=\mu^{\alpha\theta}=\mu^{\theta\beta}=e\qquad \alpha,\beta\neq\theta,\nu,\\
    &\mu^{\alpha\beta}=f\qquad\alpha,\beta\neq\theta,\nu,
    \end{split}
\end{equation}
where $\theta$ and $\nu$ are the two specific replica indices.
From $\bb{H}\bm{\mu}_3=\lambda_3\bm{\mu}_3$, we obtain
\begin{equation}
    \label{eq:18}Pd+2(n-2)Qe+\frac{1}{2}(n-2)(n-3)Rf=\lambda_3 d.
\end{equation}
The orthogonality property is given by
\begin{subequations}
\begin{align}
    \label{eq:19}&da+2(n-2)ea+\frac{1}{2}(n-2)(n-3)fa=0,\\
    \label{eq:20}&db+(n-2)eb+(n-2)ec+\frac{1}{2}(n-2)(n-3)fc=0.
    \end{align}
\end{subequations}
Using \reff{eq:18}, \reff{eq:19} and \reff{eq:20} and setting $n\to 0$, we get the $\frac{n(n-3)}{2}$-fold degenerate eigenvalue
\begin{equation}
    \lambda_3 = P-2Q+R.
\end{equation}
The total degeneracy (the number of linearly independent eigenvectors)
of these three types of eigenvectors is $n(n-1)/2$, which implies that we have exhausted all the eigenvalues.

%\bibliography{ref}

\end{document}